\documentstyle[aps,epsfig]{revtex}

\begin{document}
\draft      % \draft command makes pacs numbers print

%
%%%%%%%%%%%%%%%%%%%%%%%
% Define new commands %
%%%%%%%%%%%%%%%%%%%%%%%
%
\newcommand{\txt}[1]{{\mathrm #1}} %facilitates sub/superscripts
\newcommand{\note}[1]{{\small\tt $\left[ \right.$ #1 $\left. \right]$}}

\newcommand{\order}{\mbox{${\cal O}$}}
\newcommand{\nbody}{n~body}
\newcommand{\nplusbody}{n+1~body}
\newcommand{\BHO}{BHO}
\newcommand{\onejet}{1~jet}
\newcommand{\nojet}{0~jet}

\newcommand{\deltaS}{\delta_\txt{s}}
\newcommand{\deltaC}{\delta_\txt{c}}
\newcommand{\alphaS}{\mbox{$\alpha_\txt{S}$}}
\newcommand{\alphaEM}{ \mbox{ $\alpha_\txt{EM}$ } }
\newcommand{\NLOa}{\mbox{NLO(\alphaS)}}

\newcommand{\kT}{k^T}
\newcommand{\veckT}{\vec{k}^T}
\newcommand{\pT}[1]{P^T_{\mathrm #1}}
\newcommand{\pL}[1]{P^L_{\mathrm #1}}
\newcommand{\pX}[1]{P^x_{\mathrm #1}}
\newcommand{\pY}[1]{P^y_{\mathrm #1}}
\newcommand{\pZ}[1]{P^z_{\mathrm #1}}
\newcommand{\vecpT}[1]{\vec{P}^T_{\mathrm #1}}
\newcommand{\Minv}[1]{M^\txt{inv}(\mathrm #1)}  %Invariant mass

%
%%%%%%%%%%%%%%%%%%%%%%%%%%%%%%%%%%%%%%%%%%%%%%%%%%%%%%%%%%%%%%%%%%%%%%%%%%
%

\title{Unweighted event generation in hadronic $WZ$ production 
  at order(\alphaS)}

\author{Matt Dobbs\footnote{Electronic Mail Address: Matt.Dobbs@CERN.ch} 
  and Michel Lefebvre\footnote{Electronic Mail Address: Lefebvre@UVic.ca} }
\address{
  Department of Physics and Astronomy, University of Victoria,
  P.O.\ Box 3055, Victoria, British Columbia, Canada V8W~3P6}
\date{\today}
\maketitle

\begin{abstract}
We present an algorithm for unweighted event generation in the
partonic process $pp \rightarrow W^\pm Z (j)$ with leptonic decays at
next-to-leading order in \alphaS. Monte Carlo programs for processes
such as this frequently generate events with negative weights in
certain regions of phase space. For simulations of experimental data
one would like to have unweighted events only.  We demonstrate how the
phase space from the matrix elements can be combined to achieve
unweighted event generation using a second stage Monte Carlo
integration over a volume of real emissions (jets). Observable
quantities are kept fixed in the laboratory frame throughout the
integration. The algorithm is applicable to a broader class of
processes and is CPU intensive.
\end{abstract}

% insert suggested PACS numbers in braces on next line

%
%%%%%%%%%%%%%%%%%%%%%%%%%%%%%%%%%%%%%%%%%%%%%%%%%%%%%%%%%%%%%%%%%%%%%%%%%%
%

\section{INTRODUCTION} \label{introduction}

Next-to-leading order in \alphaS\ (\NLOa) corrections in di-boson
production are large at the CERN LHC and sizeable at the Fermilab
Tevatron (see for example \cite{EW_LHC_WORKSHOP,Baur:2000xd}).  Run~I
physics analyses (i.e., \cite{Abachi:1997xe,Abbott:1999ec}) at the
Tevatron have employed constant $k$-factors to approximate \NLOa\
corrections.  With the increased energy and luminosity of Run~II and
the new energy regime that will be probed with the CERN LHC,
simulations incorporating higher order corrections will be of
increased importance.

\NLOa\ differential cross-section predictions for di-boson production
at hadron colliders are normally accomplished with Monte Carlo
integration programs.  Divergences inherent in the Feynman graphs
contributing at \order(\alphaS) are handled with either the phase
space slicing method~\cite{Baer:1989jg} or the subtraction
method~\cite{Ellis:1981wv} with the effect that both positive weighted
events and negative weighted (or probability) events are
generated. The contribution from negative weighted events cancels in
an integration over a suitably large volume of hadronic final states,
and a physical (positive probability) prediction is
obtained. Integration programs are successful in producing
distribution and cross-section predictions for inclusive observables
subject to any experimental cuts providing statistics are large enough
to effect the cancellations, and a sufficient volume of jets has been
integrated within each histogram bin. However, owing to the negative 
probability events, unweighted event generation --- the generation of 
events with the distribution predicted by theory --- with
these \NLOa\ matrix elements has not been achieved to date. 

Unweighted event generation is often preferred over integrated
distributions because it can genuinely simulate experimental data and
because it is difficult to perform detector simulation on the large
event statistics required for integrated distributions.

Unweighted event generation can be reduced to a problem of rendering
all event weights positive definite.  If a sample of positive definite
weighted events can be obtained, the Monte Carlo hit-and-miss method
is commonly used for unweighting.  The maximum event weight
$d\sigma_{\mathrm max}$ for the process is estimated, for instance, by
sampling the cross-section a number of times.  The ratio of event
weight over the maximum event weight $\frac{d\sigma}{d\sigma_{\mathrm
max}}$ is compared to a random number $g$ generated uniformly in the
interval (0,1). Events for which the ratio exceeds the random number
($\frac{d\sigma}{d\sigma_{\mathrm max}} >g$) are accepted, the others
are rejected. This ensures the accepted events are distributed
according to the description provided by the matrix element, and so
all have unit weight.

In this paper we describe an algorithm which extends the usefulness of
the present \NLOa\ hadronic di-boson production programs by providing 
a means for unweighted event generation thereby
presenting the \NLOa\ predictions
in a format better suited for experimental simulations. It deals 
specifically with the region of low transverse momentum of the $WZ$
system, where higher order corrections are largest.  As for the
Drell-Yan process, resummation (first proposed by
\cite{Dokshitzer:1980hw}) is expected to give more reliable
predictions in this region --- however a resummed treatment of the
$WZ(j)$ process does not exist and a comparison with fixed order
predictions will always be of interest. Another approach for
this problematic region is the matching of the zero-order matrix
elements using a parton shower model for simulations of the
initial-state shower to the first order tree level (i.e.,
\order(\alphaS)) matrix elements. This has been accomplished
for the single vector boson case~\cite{Miu:1999ju,Corcella:2000gs}.

The algorithm makes use of the matrix elements from the Baur, Han, and
Ohnemus (\BHO)~\cite{BHOWZ} integration package which employs the
phase space slicing method as regularization scheme. This scheme makes
use of approximations for the analytic integration of a volume of soft
and collinear real emissions.  The approximations break
down before the volume can be increased sufficiently to render all
event weights positive.

The algorithm presented here is an extension of the phase space
slicing method.  A cylinder of jets defined by a transverse momentum
cutoff $\pT{cutoff}$ is used to partition event generation into a
subprocess where the real emission has sufficient transverse momentum
to be potentially observable (called \onejet) and another subprocess
(called \nojet) where the real emission (if present) has sufficiently
small transverse momentum such that it is taken as unobservable.
There are no divergences in the \onejet\ cross-section so these events
all have positive weights and unweighted event generation may be
accomplished by applying Monte Carlo hit-and-miss. Difficulties arise
in the treatment of the \nojet\ events, which is the topic of this
paper.

The missing transverse momentum in \nojet\ events may arise from the
invisible neutrino and from jets which escape detection. By fixing the
total missing transverse momentum rather than the individual
contributions from the invisible neutrino and unobserved jets, an
integration volume of real emissions phase space can be generated and
so the cancellations between the real emissions and one-loop graphs
can be accomplished by a Monte Carlo integration over this jet volume.
The charged lepton and missing transverse momentum vectors (i.e., the
observable quantities) are kept fixed in the laboratory frame
throughout the integration.  Since the jet volume may be large, the
fixing of the observable quantities is essential. An integration which
instead keeps other quantities such as the subprocess energy and boost
fixed (as would be the case if the phase space slicing method were
applicable to such large jet volumes) would also necessarily be an
integration over observationally different charged lepton vectors.

A final state neutrino is not the only means of generating the missing
transverse momentum over which a jet volume can be
integrated. Anything which contributes to the missing transverse
momentum may be used including the primordial $\kT$ distribution of
the proton, multiple interactions, minimum bias events, and even
detector resolution. For $WZ$ production with leptonic decays, the
invisible neutrino is a convenient choice. In generalizing this
algorithm to other processes different contributions to the missing
transverse momentum could be used.

Section~\ref{algorithm} of this paper briefly enumerates the presently
available integration packages for hadronic $WZ$ production and
outlines the regularization schemes commonly used for the matrix
elements before describing the algorithm in some detail. In
Section~\ref{results} cross-sections and several differential
cross-section distributions from integration packages are compared to
events generated with this algorithm. Our conclusions are presented in
the final section.

%
%%%%%%%%%%%%%%%%%%%%%%%%%%%%%%%%%%%%%%%%%%%%%%%%%%%%%%%%%%%%%%%%%%%%%%%%%%
%

\section{ALGORITHM FOR UNWEIGHTED EVENT GENERATION} \label{algorithm}

We choose the specific process $pp\rightarrow WZ(j)$ as a case study to
apply the algorithm.  LHC energy is chosen because the challenges in
terms of negative probability events are more severe at 14~TeV than
for the Tevatron case.

\subsection{Order \alphaS\ cross-section}

Hadronic $WZ$ production at \NLOa\ has been calculated in
Ref.~\cite{BHOWZ,Frixione:1992pj,Campbell:1999ah,Dixon:1998py}.  All
calculations with the exception of Ref.~\cite{Campbell:1999ah} use the
narrow width approximation, wherein the gauge bosons are taken on
their mass shell.
%wherein the Breit-Wigner describing the vector boson masses is 
%replaced by a delta function(i.e., on shell)
%with normalization fixed by the integral of the Breit-Wigner.
The correlations in the decays of the vector bosons to massless
fermions are fully included in
Ref.~\cite{Campbell:1999ah,Dixon:1998py}, while they are included in
all but the one-loop graphs in Ref.~\cite{BHOWZ}. For the numerical
implementation of the \NLOa\ matrix elements
Ref.~\cite{Frixione:1992pj,Campbell:1999ah,Dixon:1998py} uses the
subtraction method, while Ref.~\cite{BHOWZ} employs the phase space
slicing method. The calculations of
Ref.~\cite{Frixione:1992pj,Campbell:1999ah,Dixon:1998py} were found to
be in agreement in Ref.~\cite{Dixon:1998py}. An updated version of the
Ref.~\cite{BHOWZ} program incorporating an omitted region of phase
space has been compared with Ref.~\cite{Dixon:1998py} and found to be
in good agreement~\cite[Sec.~5.5]{EW_LHC_WORKSHOP}.

The \NLOa\ cross-section receives contributions from the square of
the Born level graphs, the interference of the Born graphs with the
one-loop graphs, and the square of the real emission graphs,
\begin{equation}
        {\cal{M}}_\txt{NLO}^2 ~=~ 
        {\cal{M}}_\txt{Born}^2 ~+~
        {\cal{M}}_\txt{Born} ~\otimes~ {\cal{M}}_\txt{OneLoop} ~+~
        {\cal{M}}_\txt{RealEmission}^2.
%       ~+~ \order(\alphaS^2) ~+~ \ldots
\end{equation}
Soft, collinear, and ultraviolet divergences appear when any of the
$\order(\alphaS)$ graphs are treated alone, and so a regularization
scheme is necessary to ensure the inherent cancellations.  The matrix
elements used in this study employ the phase space slicing method. A
volume of real emissions phase space (herein referred to as
\nplusbody\ phase space) is partitioned by means of soft and collinear
cutoffs\footnote{ The phase space slicing method defines the soft jet 
  region by
  \begin{displaymath}
    E_{\mathrm jet}^{\mathrm CMS}<\deltaS \frac{ \sqrt{\hat{s}} }{2}
  \end{displaymath}
  where $E_{\mathrm jet}^{\mathrm CMS}$ is the jet energy in the $WZj$
  subprocess center of mass frame, $\deltaS$ is the soft cutoff, and
  $\sqrt{\hat{s}}$ is the subprocess energy. The collinear region is
  defined by
  \begin{displaymath}
    |E_i E_j \pm \vec{p}_i \cdot \vec{p}_j| < \deltaC \hat{s}
  \end{displaymath}
  where $E_i$ and $\vec{p}_i$ are the energy and momentum of
  massless parton $i$ and $\deltaC$ is the collinear cutoff. For more
  details see Ref.~\cite{Baer:1989jg}.
}
 and incorporated analytically with the soft gluon and leading pole
approximations into the phase space of the diagrams lacking a real
emission (herein referred to as \nbody).

In the region where the soft and collinear divergences overlap,
the perturbative QCD calculation may be visualized in terms of an 
expansion in powers of 
\begin{equation} \label{ExpansionParameter}
  \frac{\alphaS}{2\pi} \left[ \ln \frac{\hat{s}}{{\pT{jet}}^2} \right]^2
\end{equation}
where \alphaS\ is the QCD coupling, $\sqrt{\hat{s}}$ is the subprocess
energy and characterizes the scale of the interaction, and $\pT{jet}$
is the transverse momentum of the order \alphaS\ real emission.
When the jet transverse momentum $\pT{jet}$ is small
relative to $\sqrt{\hat{s}}$, the function given in
Eq.~\ref{ExpansionParameter} becomes large and multi-gluon emission is
important --- this is the problem region for fixed order calculations.
We wish to integrate out a large enough $\pT{jet}$ volume such that
the function given in Eq.~\ref{ExpansionParameter} remains smaller
than unity.  For $\sqrt{\hat{s}}=1$~TeV this gives $\pT{jet}$ of order
10~GeV, which in turn corresponds nicely to the detector capabilities
at the LHC, wherein typically jets above 15~(30)~GeV will be
reconstructed at low (high) luminosity. Smaller jet integration
volumes would be necessary at lower machine energies.

\begin{figure}
% actual REVTeX width should be 8.6cm
  \mbox{\epsfig{file=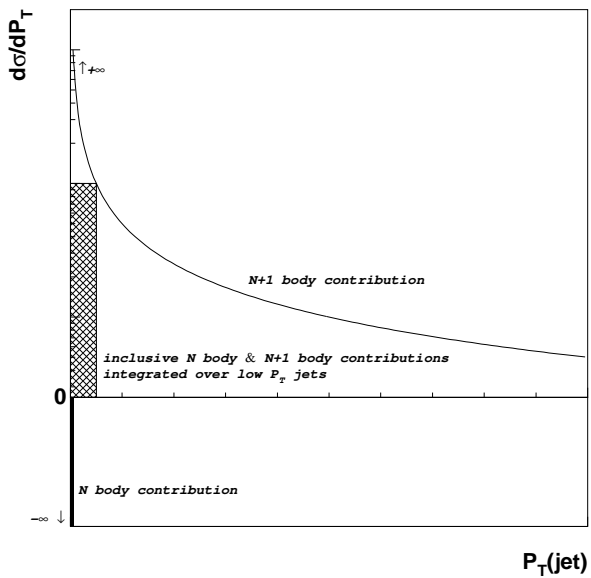,width=8.6cm}}
  \caption{
    A hypothetical transverse momentum distribution of the jet
    $\pT{jet}$ in the \NLOa\ integration is shown for illustrative
    purposes. The \nplusbody\ contribution diverges (solid curve) as
    $\pT{jet}\rightarrow0$. The \nbody\ contribution is a negative
    divergence at the origin.  The divergences cancel in the integral
    over a suitably large volume of hadronic final states (hatched
    region).}
  \label{jet_pt_partition}
\end{figure}

The situation is illustrated schematically in
Fig.~\ref{jet_pt_partition}. A hypothetical jet transverse momentum
distribution is shown for the \nbody\ and \nplusbody\ case. The
\nplusbody\ contribution diverges as the jets become soft or
collinear, while the \nbody\ contribution is a negative divergence at
the origin. The divergences cancel in an integration over a suitably
large volume of jet transverse momenta (hatched region). The phase
space slicing method effects such an integration, but the jet volume
is not sufficiently large to keep the cross-section in the inclusive
(hatched) region positive for most events.

\subsection{Generating unweighted events}

We begin by dividing events into two classes: those containing a jet
with sufficient transverse momentum ($\pT{jet}>\pT{cutoff}$) to be
classified as potentially observable (referred to here as \onejet);
and those without a jet or containing a jet with transverse momentum
below the cutoff $\pT{jet}<\pT{cutoff}$ (referred to as \nojet).  
This differs from the
definition of \nbody\ and \nplusbody\ events where the soft and
collinear cutoffs define the partition.
% Here the \nojet\ and \onejet\ events are partitioned by the 
% jet transverse momentum cutoff $\pT{cutoff}$.
In the \onejet\ region, the matrix elements are order \alphaS, but the
calculation is leading order. There are no divergences nor are there
negative weighted events, so event generation can be accomplished with
the usual Monte Carlo techniques.

The algorithm concerns events in the \nojet\ region only and we focus
exclusively on events of this type for the remainder of this paper.

The phase space slicing method is extended by integrating out the
cylinder of jets defined by the $\pT{cutoff}$ using a second stage
Monte Carlo integration. As the $\pT{cutoff}$ is chosen to be the
order of 10~GeV, there is a danger that the integration may affect the
kinematics of the event to the point which is observable by the
detector. With this in mind, the observable vectors (charged leptons
for the case of $WZ$ production with leptonic decays) are kept fixed
throughout the integration, and the center of mass frame vectors are
allowed to fluctuate.

We start by sampling the observable quantities of the event
i.e., fixing the vectors
\begin{equation}\label{FixedVectors} 
  P_{\mathrm l^\pm_W},P_{\mathrm l^+_Z},P_{\mathrm l^-_Z},\vecpT{miss}
\end{equation} 
where $P_{\mathrm l^\pm_W}$ is the momentum of the 
charged lepton from the $W^\pm$ 
decay, $P_{\mathrm l^+_Z},P_{\mathrm l^-_Z}$ are the momenta of the
charged leptons 
from the $Z$ decay, and $\vecpT{miss}$ is the missing transverse 
momentum. With the $Z$ boson taken on shell and requiring no net
momentum transverse to the beam, this amounts to 8 degrees of freedom.

For the case of the \nbody\ matrix element these vectors also fix the
neutrino four-vector with $\vecpT{\nu}=\vecpT{miss}$ and the $W$-mass
constraint gives zero or two solutions for the neutrino longitudinal
momentum. Within the \BHO\ matrix elements a further 4~degrees of
freedom specify the \NLOa\ corrections. Thus the
average \nbody\ weight $\left< d\sigma^{\mathrm \nbody} \right>$ for the
observable event is arrived at
by integrating over these degrees of freedom and averaging over the
two neutrino solutions. This weight is usually negative. When no
neutrino solutions exist the weight is zero.  We refer to a point in
the full \nbody\ phase space (defined by the vectors of
Eq.~\ref{FixedVectors}, the 4 extra degrees of freedom, and the
neutrino solution choice) as a ``subevent'', such that an ``event'' as
defined by this algorithm is an integration over many
subevents.\footnote{ A subevent from this algorithm corresponds
  precisely to an event from a Monte Carlo integration package.}

We now incorporate the cylinder of jets defined by the $\pT{cutoff}$
by performing a similar integration over \nplusbody\ subevents. For
each subevent a jet with transverse momentum up to the cutoff is
sampled (3 degrees of freedom). The jet transverse momentum fixes the
invisible neutrino transverse momentum for the event with the
constraint that the observed missing transverse momentum is kept
fixed,
\begin{equation}
  \vecpT{\nu} = \vecpT{miss} - \vecpT{jet}.
\end{equation}
The longitudinal neutrino momentum is specified up to a two-fold ambiguity
by the $W$-mass constraint. The charged lepton vectors are also kept
fixed in the laboratory frame, such that all the quantities in
Eq.~\ref{FixedVectors} are unchanged throughout the integration.
The average subevent weight is $\left< d\sigma^{\mathrm \nplusbody} \right>$,
which is a positive number.

The phase space for a Monte Carlo integration is normally sampled in
terms of variables such as the invariant masses, rapidities, and decay
angles which give a triangular transformation matrix such that the
Jacobian is the product of the diagonal entries (as is the case for
the di-boson integration packages).  The fixed charged lepton vectors
in the laboratory frame constraint introduces inherent correlations,
and the transformation matrix becomes more complicated: for the $WZ$
\onejet\ case it is a $11\times11$ matrix with no zero elements.
Fortunately it is simple to calculate this matrix and its Jacobian
numerically on an event by event basis, which is the strategy we have
used. This Jacobian is included in the subevent weights
$d\sigma^{\mathrm \nbody}$ and $d\sigma^{\mathrm \nplusbody}$.

The total event weight for the configuration defined by
Eq.~\ref{FixedVectors} is the sum of the \nbody\ and \nplusbody\
contributions, with a factor which accounts for the observable vectors
phase space sampling
\begin{equation}\label{EventWeight}
d\sigma^\txt{\nojet}~=~
        W_\txt{PhaseSpace} \times \left(~
                \left<d\sigma^{\mathrm \nbody}\right> 
            ~+~ \left<d\sigma^{\mathrm \nplusbody}\right> 
        ~\right).
\end{equation}

It is unimportant that the number of subevents for the second stage
integration be sufficiently large to give an accurate result for each
observable event. A poor accuracy simply means two identically
configured events may have differing event weights, however by grace
of the Monte Carlo method, the average converges to give the correct
cross-section. The subevent statistics are required to be sufficiently
large to ensure the event weights of Eq.~\ref{EventWeight} very
rarely fall negative on account of statistical fluctuations. We refer
to events where this occurs as ``remnant negative weight events''.
Events of this type are discarded, and thus contribute to a (biased)
systematic Monte Carlo error. The number of subevents is a means of
directly controlling this error. It is easy to evaluate this error.

Finally, we note the application of this algorithm to the re-weighting
of leading order events (i.e., re-scaling of tree level event weights
to account for \NLOa\ corrections) needs to be approached with
caution, as the phase space volume of observable vectors in
Eq.~\ref{FixedVectors} is larger for the \nplusbody\ case than
for the \nbody\ case since only limited configurations of these
vectors provide \nbody\ neutrino longitudinal momentum solutions
which satisfy the $W$-mass constraint.  This difference needs to be
accounted for in a re-weighting scheme.

%
%%%%%%%%%%%%%%%%%%%%%%%%%%%%%%%%%%%%%%%%%%%%%%%%%%%%%%%%%%%%%%%%%%%%%%%%%%
%

\section{RESULTS AND COMPARISON WITH INTEGRATION PACKAGES} \label{results}

The results presented in this section use identical parameters as the
comparison of integration packages in
Ref.~\cite[Sec.~5.5]{EW_LHC_WORKSHOP} with the exception of the jet
definition for the jet veto. In that study the jet transverse momentum
cutoff $\pT{cutoff}$ was motivated by what the detectors are capable
of measuring. Here we motivate the $\pT{cutoff}$ by what the detector
is incapable of reconstructing: a $\pT{cutoff}$ of 15~GeV is defined
to partition the \onejet\ and \nojet\ regions (there is no jet
rapidity requirement).  The {\tt CTEQ4M}~\cite{Lai:1997mg} structure
functions are used.  Input parameters are taken as
$\alpha_{EM}(M_Z)=\frac{1}{128}$, $\sin^2\theta_W=0.23$,
$\alpha_s(M_Z)=0.1116$, $M_W=80.396$~GeV, $M_Z=91.187$~GeV,
factorisation scale $Q^2=M_W^2$, and Cabibbo angle
$\cos\theta_\txt{Cabibbo} =0.975$ with no 3rd generation mixing.
Branching ratios are taken as Br($Z\rightarrow l^+ l^-$) = 3.36\%,
Br($W^\pm\rightarrow l^\pm\nu$)=10.8\%.  The $b$ quark contribution to
parton distributions has been taken as zero.  Kinematic cuts motivated
by anomalous triple gauge boson coupling analyses are chosen (see for
example \cite[Sec.~5]{EW_LHC_WORKSHOP}). The
transverse momentum of all charged leptons must exceed 25~GeV and the
rapidity of all charged leptons must be less than 3 in magnitude. Missing
transverse momentum must be greater than 25~GeV.  The \BHO\ package is
employed for the evaluation of matrix elements.

We first evaluate the algorithm's ability to integrate out the
negative probability events. The second stage integration is
accomplished with relatively low statistics to keep the computer
processing time reasonable, only 100 subevents of each type (\nbody\
and \nplusbody) are used.

For the default choices of the soft and collinear cutoffs in the \BHO\
matrix elements ($\deltaS=0.01,~\deltaC=0.001$) the cancellations
between the \nbody\ and \nplusbody\ partitions are very large, and the
second stage integration would need unreasonably large statistics for
good behavior.\footnote{The same problematic behavior is observed when matrix
elements employing the subtraction method are used. We found no
immediate solution to this problem for the subtraction method, and so
we have been unable to use this algorithm with subtraction method
matrix elements thus far.} In order to
achieve a portion of the jet volume integration analytically we
increase the cutoffs ($\deltaS=0.05,~\deltaC=0.002$) which is well
within the range of applicability of the soft gluon and leading pole
approximations. We emphasize that experiments must take the
integration volume into account by ensuring that their analysis
is not sensitive to jets within the volume (as defined by the jet
momentum cutoff of this algorithm {\it and} the soft and collinear
cutoffs of the phase space slicing method). This consideration is true
in general for the phase space slicing method.

With these choices of the phase space slicing cutoffs ``remnant
negative weight events'' account for approximately 0.1\% of the total
sample of weighted events and -0.02\% of the cross-section.  Most of
these events are the result of the low second stage integration
statistics such that re-evaluating these events with a larger number
of subevents renders the event weight positive.

There is another class of remnant negative events which are not due to 
statistics. For these events the charged lepton in the $W$-decay is 
emitted along the line of flight of the $W$ in its rest frame.  This
configuration is disfavored by the spin correlations, wherein the
$W^+$-decay ($W^-$-decay) with a left-handed (right-handed) particle
emitted along the $W$ line of flight is preferred.  The \BHO\ matrix
elements use a spin averaged treatment for the one-loop graphs and so
these correlations are not at all included for the virtual
corrections (though spin correlations are included everywhere
else).  This means events in this configuration receive an
unphysically large negative contribution from the one-loop graphs
while receiving the correct (spin correlation suppressed) contribution
from the other graphs.  Thus this class of negative event weights
arise as a result of the lepton correlation approximations used in the
\BHO\ matrix elements. The algorithm is sensitive to these
approximations and would benefit from improved theoretical
calculations (as are included
in~\cite{Campbell:1999ah,Dixon:1998py}). The negative weights
associated to events of this type are extremely small.

Finally there are a few events with negative weights at extraordinary
subprocess energy ($\sqrt{\hat{s}}>5$~TeV) where the total
cross-section is orders of magnitude below one event per LHC
year. These events may arise from a combination of the lepton
correlation effect and an insufficiently large jet volume for events
at this energy scale, as suggested by the function given in 
Eq.~\ref{ExpansionParameter}.

\subsection{Comparison with the \BHO\ integration package}

In Table~\ref{xsecComparison} we enumerate the leading order and
\NLOa\ predictions for \nojet\ $WZ$ production from the \BHO\
integration package together with the result from this algorithm which
employs the \BHO\ matrix elements. The \NLOa\ prediction differs
considerably from the leading order one and this algorithm is able
to reproduce the \NLOa~\BHO\ predictions well.  The accuracy of the
prediction from this algorithm suffers slightly with respect to that
from the \BHO\ integration package alone in that the correlations
between integration variables make it more difficult for the adaptive
integration to optimize the phase space sampling.

It takes considerably more computer time to evaluate the \NLOa\ matrix
elements than for the leading-order case. This algorithm requires a
further factor 200 over the \NLOa\ integration package in evaluating
one weighted event because of the 2$\times$100 subevents in the second
stage integration. 

Monte Carlo hit-and-miss is used to extract a sample of unweighted
events from the (positive definite) events generated using this
algorithm. No attempt has been made to optimize for efficiency, and
the time required for the generation of one unweighted event is quite
large\footnote{ 200~seconds are required for one unweighted event
        using a 600~MHz personal
        computer running Linux 2.2. Pythia~\cite{Sjostrand:1994yb} 
        requires only about 0.02~seconds for the generation of one 
        leading order $WZ$ event with an efficiency of 1-2\%.  
} with an efficiency of about
0.1\%. Nevertheless it is possible to generate an event sample of a
few low luminosity LHC years (i.e., several thousand events) in
several hours using computer farms of the size currently in use at
most universities or experiments.

\begin{figure}
  \mbox{\epsfig{file=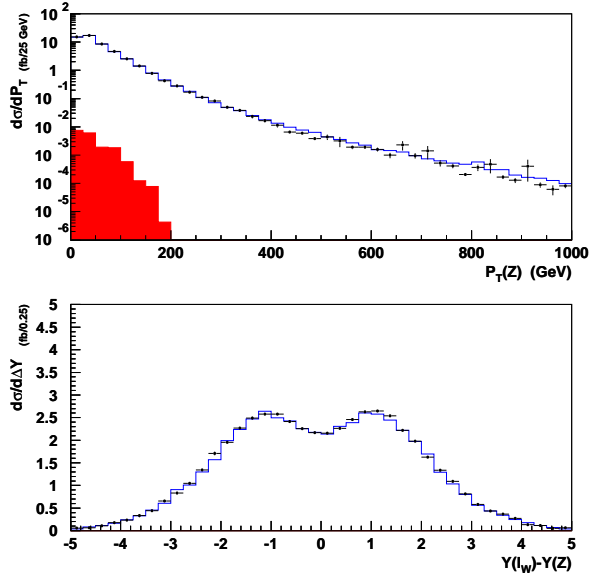,width=8.6cm}}
  \caption{
    The transverse momentum distribution of the $Z$ ($\pT{Z}$, top)
    and the rapidity separation between the $Z$ and the lepton from
    the $W$ decay ($Y(l_W)-Y(Z)$, bottom) are compared for weighted
    events from the \BHO~\NLOa\ generator (solid open histogram,
    statistical errors are small) and for weighted events from the
    algorithm (points with error bars). Only positive weighted events are
    included in the black algorithm distribution, and the errors are
    statistical only.  
    The distribution of discarded negative weighted events from the
    algorithm are shown in the lower shaded histogram (using the
    absolute value of the weights).  This distribution is several
    orders of magnitude below the positive weight event distribution,
    and is only observable on the logarithmic upper plot.
    The distributions are for the \nojet\ case with kinematic cuts
    as described in the text.
    }
  \label{compare_weighted}
\end{figure}

\begin{figure}
  \mbox{\epsfig{file=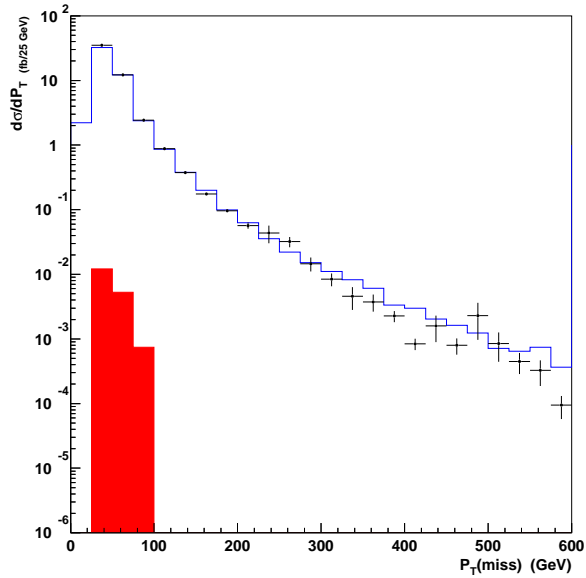,width=8.6cm}}
  \caption{
    The missing transverse momentum distribution $\pT{miss}$ is
    compared for weighted events from the \BHO~\NLOa\ generator (solid
    open histogram, statistical errors are small) 
    and for weighted events from the algorithm (points
    with error bars).  Only positive weighted events are included in
    the black algorithm distribution, and the errors are statistical
    only.
    The distribution of discarded negative weighted events from the
    algorithm are shown in the lower shaded histogram (using the
    absolute value of the weights).  This distribution is several
    orders of magnitude below the positive weight event distribution.
    The distribution is for the \nojet\ case with kinematic cuts
    as described in the text.}
  \label{compare_weighted_ptmiss}
\end{figure}

Weighted \nojet\ event distributions are compared in
Fig.~\ref{compare_weighted}~\&~\ref{compare_weighted_ptmiss} for the
\BHO\ integration package and this algorithm.
Figure~\ref{compare_weighted}(top) shows the transverse momentum
distribution of the $Z$, which is the distribution normally used to
probe anomalous $WWZ$ couplings at hadron
colliders~\cite{Zeppenfeld:1988ip}. The rapidity separation between
the $Z$-boson and the charged lepton from the $W$-decay
(Fig.~\ref{compare_weighted}, bottom) is sensitive to the approximate
radiation zero in hadronic $WZ$ production~\cite{Baur:1994ia}. Finally
the missing transverse momentum distribution is shown in
Fig.~\ref{compare_weighted_ptmiss}. The jet integration volume arises
from this distribution, and so it is the place where discrepancies
between this algorithm and the integration package might arise. The
agreement between the \BHO\ integration package and this algorithm in
all of these distributions is good.

In order to evaluate the effect of remnant negative weight events, the
absolute value of the weights of negative events is plotted in
Fig.~\ref{compare_weighted}~\&~\ref{compare_weighted_ptmiss}
(shaded regions). The remnant negative weight event distribution is
several orders of magnitude below the positive event distribution in
all cases and is largest in the region where the differential
cross-section is also largest, and so the effect of these events is
negligible everywhere.  Histograms of this sort provide the means of
evaluating the effect of these remnant negative events. For faster
event generation the second stage integration statistics may be
decreased, resulting in an increase in the size of this shaded
region. It is impossible to completely eliminate the shaded region of
remnant negative events by increasing the second stage integration
statistics indefinitely, due to the lepton correlation effects
discussed in the first part of this section. If the negative weighted
events from the \BHO\ integration package (without this algorithm)
were also superimposed, they would be distributed uniformly about half
an order of magnitude below the \BHO\ prediction.

\begin{figure}
  \mbox{\epsfig{file=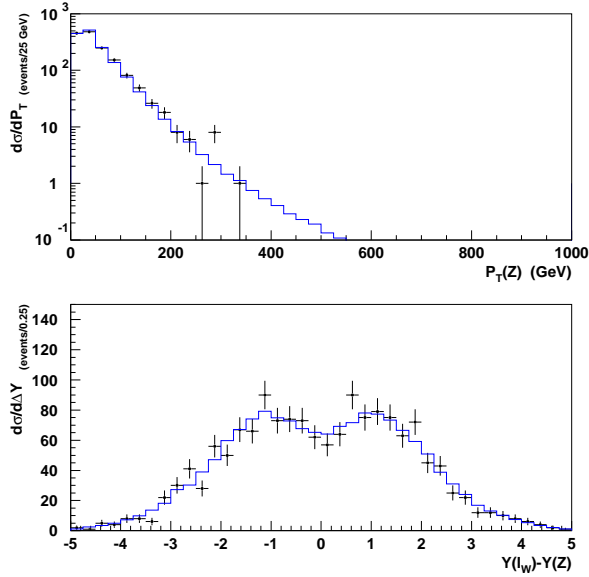,width=8.6cm}}
  \caption{
    The transverse momentum distribution of the $Z$ ($\pT{Z}$, top)
    and the rapidity separation between the $Z$ and the lepton from
    the $W$ decay ($Y(l_W)-Y(Z)$, bottom) are compared for weighted
    events from the \BHO~\NLOa\ generator (solid open histogram,
    statistical errors are small) and for unweighted events from the
    algorithm (points with error bars).  The distributions are for an
    integrated luminosity corresponding to three years of low
    luminosity running at the LHC (30~fb$^{-1}$). The error bars
    assume Poisson statistics.
    The distributions are for the \nojet\ case with kinematic cuts
    as described in the text.}
  \label{compare_unweighted}
\end{figure}

In Fig.~\ref{compare_unweighted} unweighted events from this
algorithm are superimposed on the \BHO\ integration package prediction
for the same distributions as Fig.~\ref{compare_weighted}. The error
bars assume Poisson statistics and the agreement is good. This figure
represents the first unweighted event generation at \NLOa\ for the
partonic process $pp\rightarrow WZ(j)$.

%
%%%%%%%%%%%%%%%%%%%%%%%%%%%%%%%%%%%%%%%%%%%%%%%%%%%%%%%%%%%%%%%%%%%%%%%%%%
%

\section{CONCLUSIONS} \label{conclusions}

We have described an algorithm for unweighted event generation at
\NLOa\ for the partonic process $pp\rightarrow WZ(j)$ with leptonic
decays and evaluated its effectiveness at LHC energy. 
Event generation consists of a
two stage Monte Carlo integration, and so requires considerable
computer time. The possibility of optimizing the generation and the
code itself has not been addressed, and certainly the performance may
be considerably improved. The partition of events into \nojet\ and
\onejet\ regions is motivated by experimental capabilities and 
defined by a jet transverse momentum cutoff. This
partition, as well as the \nbody\ and \nplusbody\ partition inherent
in the phase space slicing method, must be taken into account in
experimental analyses.

Generation of high transverse momentum real emissions is not addressed
by this algorithm, but this region of events is free of divergences
and so unweighted event generation is easily achieved with the usual
Monte Carlo methods.

A small number of remnant negative weighted events are discarded by
the algorithm. The effect of these events is negligible and
contributes to a biased systematic Monte Carlo error which may be
monitored and evaluated.

The agreement between events generated with this algorithm and the 
\BHO~\NLOa\ integration package is good.

The algorithm is generalizable to a broader class of
processes. Hadronic $W\gamma (j)$ production can be treated in an
identical way, while the $W^+W^- (j)$ process with leptonic decays can
be handled by treating the neutrino (anti-neutrino) as invisible
and the anti-neutrino (neutrino) as an observable lepton for the
purposes of this algorithm.  Other means of generating the jet
integration volume besides the invisible neutrino may be used for
other processes.

%
%%%%%%%%%%%%%%%%%%%%%%%%%%%%%%%%%%%%%%%%%%%%%%%%%%%%%%%%%%%%%%%%%%%%%%%%%%
%

\section*{ACKNOWLEDGMENTS} \label{acknowledgments}

The authors would like to thank the ATLAS collaboration which
motivated this project.
We are grateful to Ulrich Baur for providing the matrix elements as well as
informative correspondence and discussions. We thank Ian Hinchliffe
and J$\o$rgen Beck Hansen for useful comments.
This work has been supported by the Natural Sciences and
Engineering Research Council of Canada.

%
%%%%%%%%%%%%%%%%%%%%%%%%%%%%%%%%%%%%%%%%%%%%%%%%%%%%%%%%%%%%%%%%%%%%%%%%%%
%

%
%%%%%%%%%%%%%%%%%%%%%%%%%%%%%%%%%%%%%%%%%%%%%%%%%%%%%%%%%%%%%%%%%%%%%%%%%%
%

\section*{FIGURES}

%
%%%%%%%%%%%%%%%%%%%%%%%%%%%%%%%%%%%%%%%%%%%%%%%%%%%%%%%%%%%%%%%%%%%%%%%%%%
%

%
%%%%%%%%%%%%%%%%%%%%%%%%%%%%%%%%%%%%%%%%%%%%%%%%%%%%%%%%%%%%%%%%%%%%%%%%%%
%

%
%%%%%%%%%%%%%%%%%%%%%%%%%%%%%%%%%%%%%%%%%%%%%%%%%%%%%%%%%%%%%%%%%%%%%%%%%%
%

%
%%%%%%%%%%%%%%%%%%%%%%%%%%%%%%%%%%%%%%%%%%%%%%%%%%%%%%%%%%%%%%%%%%%%%%%%%%
%

%
%%%%%%%%%%%%%%%%%%%%%%%%%%%%%%%%%%%%%%%%%%%%%%%%%%%%%%%%%%%%%%%%%%%%%%%%%%
%

\section*{TABLES}

\begin{table}
\caption{
The integrated cross-section subject to the cuts described in the text 
is compared for the leading order \BHO\ calculation, the \NLOa\ \BHO\ 
calculation, and for this algorithm which uses the \BHO\ matrix elements.
}
\label{xsecComparison}
\begin{tabular}{lccc}
  \multicolumn{2}{c}{ \nojet\ LHC $W^+Z$ production cross-section} \\ \hline
    Baur/Han/Ohnemus LO     & $70.5\pm0.1$~fb \\
    Baur/Han/Ohnemus \NLOa  & $51.0\pm0.1$~fb \\
    This 2 Stage Algorithm  & $50.5\pm0.3$~fb \\
\end{tabular}
\end{table}

%
%%%%%%%%%%%%%%%%%%%%%%%%%%%%%%%%%%%%%%%%%%%%%%%%%%%%%%%%%%%%%%%%%%%%%%%%%%
%

\end{document}